\documentclass[12pt,a4paper]{article}
\usepackage{xcolor}
\usepackage{amsmath,amssymb,amsthm,amsfonts}
\usepackage[margin=1.0in,rmargin=1.in,tmargin=2cm,bmargin=3cm]{geometry}
\usepackage{cite}
\usepackage{graphicx}
\usepackage{subcaption}
\usepackage{epsfig}
\usepackage{hhline}
\usepackage[pagebackref=false]{hyperref}
\usepackage[affil-it]{authblk}
\numberwithin{equation}{section}
\def\a{\alpha}
\def\b{\beta}
\def\g{\gamma}
\def\d{\delta}

\def\k{\kappa}
\def\l{\lambda}
\def\m{\mu}
\def\n{\nu}

\def\r{\rho}

\def\s{\sigma}
\def\t{\tau}

\def\Gg{{\mathcal G}}

\def\S{\Sigma}
\def\G{\Gamma}

\def\L{\Lambda}

\newcommand{\nn}{\nonumber}

\def\be{\begin{equation}}
\def\ee{\end{equation}}
\def\bea{\begin{eqnarray}}
\def\eea{\end{eqnarray}}
\def\bg{\begin{align}}
\def\eg{\end{align}}

\def\pa{\partial}

\def\td{\tilde}

\makeatletter
\renewcommand\section{\@startsection {section}{1}{\z@}%
	{-3.5ex \@plus -1ex \@minus -.2ex}
	{2.3ex \@plus.2ex}%
	{\normalfont\large\bfseries}}
\renewcommand\subsection{\@startsection{subsection}{2}{\z@}%
	{-3.25ex\@plus -1ex \@minus -.2ex}%
	{1.5ex \@plus .2ex}%
	{\normalfont\bfseries}}
\makeatother

\begin{document}
	\begin{titlepage}
		{\title{{Holographic complexity in general quadratic curvature theory of gravity}}}
		\vspace{.5cm}
		\author{Ahmad Ghodsi \thanks{a-ghodsi@ferdowsi.um.ac.ir}}
		\author{Saeed Qolibikloo \thanks{s.qolibikloo@mail.um.ac.ir}}
		\author{Saman Karimi \thanks{karimi.saman@mail.um.ac.ir}}
		\vspace{.5cm}
		\affil{Department of Physics, Faculty of Science,     
			\hspace{5.5cm}Ferdowsi University of Mashhad, 	
			\hspace{7.5cm} Mashhad, Iran}
		\renewcommand\Authands{ and }
		\maketitle
		\vspace{-12cm}
		\begin{flushright}
		\end{flushright}
		\vspace{10cm}
		\begin{abstract}
In the context of CA conjecture for holographic complexity, we study the action growth rate at late time approximation for  general quadratic curvature theory of gravity. We show how the Lloyd's bound saturates for charged and neutral black hole solutions. We observe that a second singular point may modify the action growth rate to a value other than the Lloyd's bound. Moreover, we find the universal terms that appear in the divergent part of complexity from computing the bulk and joint terms on a regulated WDW patch.
		\end{abstract}
	\end{titlepage}

\section{Introduction}\label{Introduction}
In recent years, by combining the basic ideas of Quantum Information Theory (QIT) and AdS/CFT duality, a significant development in the area of black hole physics has occurred and we have witnessed the appearance of new paradigms in our quest to understand the quantum theory of gravity.

Holographic Entanglement Entropy (HEE) is a successful example and has been captured the imagination of many people in this area of research \cite{Ryu:2006bv, Ryu:2006ef}. Meanwhile, it was claimed in \cite{Susskind:2014moa} that the HEE is not enough to distinguish the degrees of freedom inside a black hole, because even if the space-time reaches the thermal equilibrium, the volume of the black hole continues to grow. Instead, complexity is an interesting proposal that characterizes these degrees of freedom.

In QIT, complexity is defined as the minimum number of quantum gates, essential to produce a particular state  $|\psi \rangle$ from a reference state $|\psi_0\rangle$.  In other words, complexity is a measure of how hard it is to construct a final state from an initial state \cite{watrous2009quantum,2014arXiv1401.3916G,0034-4885-75-2-022001}.

In the context of AdS/CFT there are two main proposals on how to compute the complexity of a boundary state. The first one is the  “Complexity=Volume”  or (CV) conjecture \cite{Susskind:2014rva,2014arXiv1403.5695S,Stanford:2014jda} and the second one is  the “Complexity=Action”  or (CA) conjecture \cite{Brown:2015bva,Brown:2015lvg}. 
The CV duality states that the complexity of a holographic boundary state on a time slice $\Sigma$ is given by
\begin{equation}
C_V(\Sigma) =\ \mathrel{\mathop {\rm Max}_{\scriptscriptstyle{\S=\partial B}} {}\!\!}\left[\frac{V(B)}{G_N \, \ell}\right]\, ,
\end{equation}
where $\ell$ is a certain length scale of the geometry (for example the curvature scale or horizon radius) and $B$ is the corresponding bulk surface. 
The CA conjecture claims that the complexity is given by the gravitational action evaluated on the Wheeler-DeWitt (WDW) patch. This patch is defined as the domain of dependence of the Cauchy surface in the bulk which asymptotically approaches the time slice $\Sigma$ on the boundary
\begin{equation}\label{CAC}
C_A(\Sigma)=\frac{I_{\rm WDW}}{\pi \hbar}\, .
\end{equation}
In CA conjecture we are dealing with evaluating the gravitational action on various space-time regions including space/time-like boundaries, null boundary surfaces, and various types of joints at the intersection of boundaries \cite{NullBound}.

Other than complexity, there are related interesting parameters, which play major roles in the computation of CV and CA conjectures. For example complexity of formation $\Delta C$, the structure of UV divergences of the complexity, the time dependence or the rate of complexity ${dC}/{dt}$, and the so-called Lloyd's bound. In what follows, we briefly review these parameters one by one.

Let us begin by introducing the complexity of formation.
The thermofield double (TFD) state can be considered as the dual description of the full geometry of an eternal AdS black hole \cite{eternal}
\begin{equation}
| {\rm TFD} \rangle = Z^{-1/2} \sum_i e^{-{E_i}/({2T})} |E_i\rangle_L |E_i\rangle_R\,,
\label{TFD}
\end{equation}
where the corresponding asymptotic boundaries (denoted by L and R) are two copies of the same CFT. One can construct thermal density matrix of the CFT at temperature $T$, by integrating out either the left or right degrees of freedom in the above state. Moreover, the entanglement between these two sets of degrees of freedom leads to the appearance of Einstein-Rosen bridge in the bulk \cite{eternal,EPR}. The complexity of formation is the difference between complexity in the process of forming the entangled TFD state  and preparing two individual copies of the vacuum state of the left and right boundary CFTs
\begin{equation}
\Delta C = \frac{1}{\pi\hbar}\big[\, I(\text{BH}) - 2\,I(\text{AdS})\,\big]\,.
\label{Cform}
\end{equation}
In \cite{Chapman:2016hwi} it was found that for boundary dimensions greater than two, the complexity of formation grows linearly with the thermal entropy at high temperatures $\Delta C \sim S$.

The growth rate of action within the WDW patch at late time
approximation or $dC/dt$ is another important factor. One of the interesting features of this parameter is that it tends to a specific universal value.
In QIT this limit has been found in the study of an arbitrary quantum system \cite{Lloyd} (Lloyd bound). In the context of CA conjecture it was first studied in \cite{Brown:2015bva, Brown:2015lvg}, where they show that the bound is violated for large enough Schwarzschild AdS black holes, and is saturated for small ones.  
Their results suggest that the growth rate would be bounded by the total energy of the system
\begin{align}\label{eq:complexitybound}
\frac{dC}{dt}\leq\frac{2E}{\pi\hbar}\,.
\end{align}
Further exploration of this bound, in the case of rotating and charged black holes were carried out in \cite{Cai:2016xho, Huang:2016fks}, where a modified formula for the action growth rate was presented. For example in the case of charged black holes
\begin{align}\label{didt}
\frac{dI}{dt}&=( M-\mu_{+} Q)-( M-\mu_{-} Q) \,,
\end{align}
where $M$ and $Q$ are mass and charge of the black hole and $\mu$ is the chemical potential. The $\pm$ signs represent the outer and inner horizons of the black hole.

In the context of gauge/gravity duality, it is usual to consider a given classical theory of gravity with an asymptotically AdS solution as a toy model. Such a model may have not a unique dual CFT but it might probably be a truncation of many various theories with additional fields that have an explicit dual CFT. Therefore by studying the holographic dictionary we explore a family of CFTs in the same universality class.
The holographic dual of Einstein gravity is restricted to a specific set of CFTs. To study a broader spectrum of CFTs, one can consider higher curvature theories of gravity in the bulk. This enables us to explore other universality classes as well because as it is observed in these theories there are various types of degrees of freedom in addition to the Einstein modes.

In this regard, the higher order curvature theories of gravity usually have two important roles. Either they show the existence of a certain property that holds for all holographic dual CFTs, for instance, the holographic c-theorems or holographic entanglement entropy, or in opposite direction, they provide counterexamples, for example in the known Kovtun-Son-Starinets  bound for the shear viscosity over entropy density ratio, where the bound is violated in the presence of certain higher curvature terms.
These properties motivate us to study the holographic complexity in both directions in this paper. In section two we examine possible situations where the Lloyd’s bound is preserved or violated and in section three we find the universal terms that appear in the divergent part of complexity.

In fact, there has been extensive discussion on the preservation or violation of the Lloyd's bound in various gravitational setups \cite{Cai:2017sjv, Carmi:2017jqz, Goto:2018iay, Fan:2019mbp, Jiang:2019fpz, Fan:2019aoj, Jiang:2019qea, Jiang:2018pfk, Sun:2019yps}. For example in \cite{Jiang:2018pfk}, using Noether charge formalism of Iyer and Wald \footnote{see also \cite{Couch:2016exn, Fan:2018wnv}}, the authors have found a modified version of the Lloyd's bound in multiple Killing horizons black hole for a higher curvature theory of gravity.
Also, in \cite{Yang:2016awy} it was shown that strong energy condition is a sufficient condition to ensure the bound inequality \eqref{eq:complexitybound} and it was argued that the equality \eqref{didt} satisfies the bound \eqref{eq:complexitybound}.

Generally, there are two main methods for computation of the action growth rate in CA conjecture, the BRSSZ method introduced in \cite{Brown:2015lvg}, and the LMPS method \cite{NullBound}. 
In the latter paper, the authors show that the two methods lead to the same result for the action growth rate in Einstein gravity. Also in \cite{Jiang:2019pgc} it was pointed out that even in general higher curvature theories of gravity these two methods provide the same result. 

Since in the first approach one does not need to know the corresponding action of the null boundaries in general higher curvature theories of gravity, we use this method to compute the action growth rate. In the BRSSZ method for the computation of $dC/dt$, in addition to the bulk action, it is only necessary to know the action on the time/space-like boundaries, which is essentially the Gibbons-Hawking-York (GHY) action and its generalizations (see section \ref{C-dot in GQC} for more details).

For a general higher curvature gravitational theory it is hard to find an appropriate surface term to make the variational principle well-posed \cite{Dyer:2008hb}, but the non-null surface terms have been developed for some gravitational theories, such as $F(R)$, Gauss-Bonnet gravity and Lanczos-Lovelock theory  \cite{delaCruzDombriz:2009et, Guarnizo:2010xr, Bunch, Myers:1987yn, Padmanabhan:2013xyr, Bueno:2016dolw} and other higher curvature theories \cite{Teimouri:2016ulk}. Specifically for $f(Riemann)$ theories of gravity by using the auxiliary field formalism these surface terms have been investigated in \cite{Deruelle:2009zk}.
In this context, complexity has been studied for higher curvature theories of gravity of example see
\cite{Alishahiha:2017hwg, Ding:2018ibq, Jiang:2018sqj, Cano:2018aqi, An:2018dbz}. Inspired by this, we are going to consider the General Quadratic Curvature (GQC) theory of gravity in this paper and compute the action growth rate in this theory.

Another related subject is the structure of UV divergences of the complexity. In both complexity conjectures, we evaluate quantities which diverge by extending near the asymptotic boundaries. 
These divergences are related to the existence of short-scale correlations in the dual boundary CFTs. As it was shown in \cite{Carmi:2016wjl}  the coefficients in these divergent terms can be written in terms of intrinsic
and extrinsic boundary curvatures. The general structure of divergent terms in complexity in $d$ dimensional space-time can be written as
\begin{equation}\label{Cuniv}
C_A\sim\frac{a_1}{\delta^{d-2}}+\frac{a_2}{\delta^{d-4}}+\cdots+\log\left(\frac{\ell}{\delta}\right)\left[\frac{b_1}{\delta^{d-2}}+\frac{b_2}{\delta^{d-4}}+\cdots\right]\,,
\end{equation}
where  $\delta$ is the UV cut-off. Here the coefficients $a_i$ depend on the regularization scheme but the coefficients $b_i$ are  universal (regulator independent).
In section \ref{universal terms} we will study the structure of these divergences of the complexity and compute the universal terms of the complexity in GQC gravity.

\section{The action growth rate of GQC gravity} \label{C-dot in GQC}
\begin{figure}[!hbt]
	\centering
	\includegraphics[scale=1]{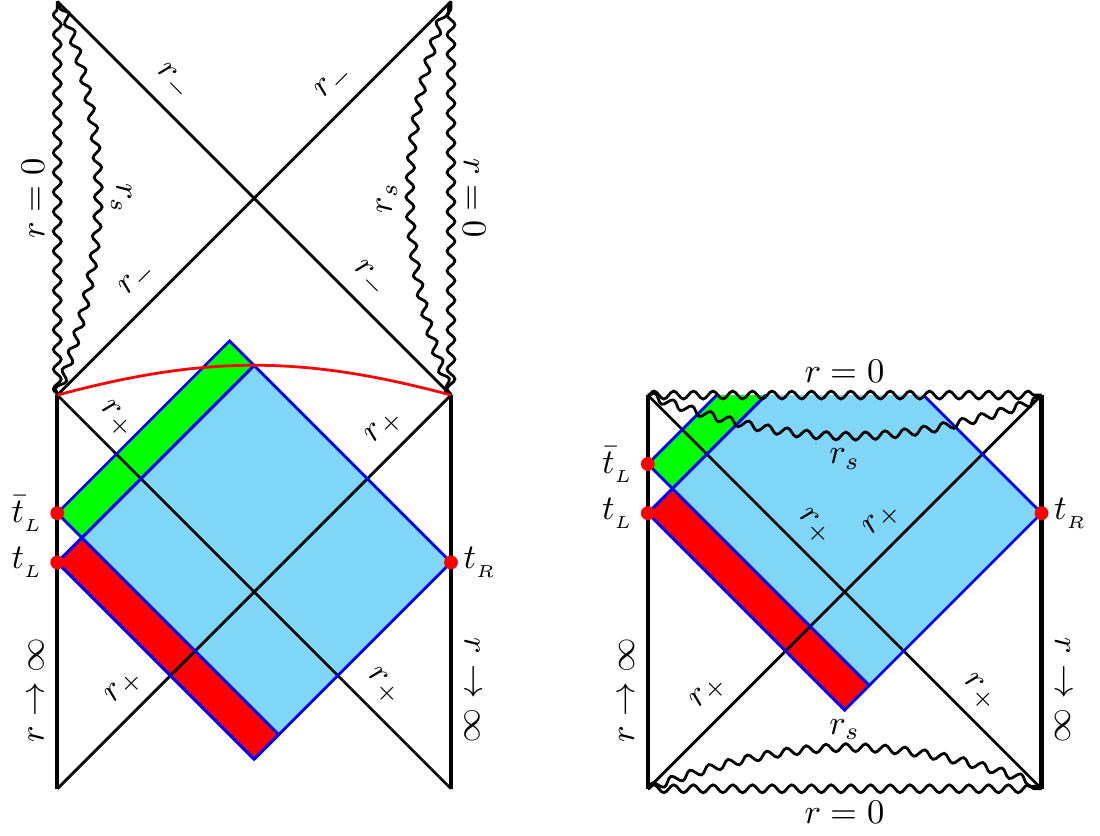}	
	\caption{\footnotesize{Penrose diagram of charged (left) and neutral (right) black holes in GQC gravity. In the charged case we have depicted the outer horizon $r_+$ and the inner horizon $r_-$, together with two singularities, one at the center of space-time $r=0$, and one behind the inner horizon, at $r=r_s$, denoted by the wiggly arcs in this picture. In the neutral case, however, the inner horizon disappears and space-like singularities appear, therefore the WDW patch should be ended on the singularity.}}
	\label{Fig1}
\end{figure}

Following \cite{Brown:2015lvg} and \cite{Cai:2016xho} we are going to compute the action growth rate on a WDW patch associated with a two-sided black hole in GQC theory of gravity, see figure \eqref{Fig1}. It is believed that this would be dual to the rate of
growth of the complexity of the boundary state. 
This black hole is a charged/neutral Schwarzschild AdS solution of the equations of motion. For charged solution (left diagram) we suppose outer and inner horizons at $r=r_\pm$. For neutral case (right diagram) we consider inside the black hole is limited from singularity to the future horizon.

According to the arguments of \cite{Brown:2015lvg}, the WDW patch is bounded by $t_L$ and $t_R$ on the left and right
for a (un)charged black hole. When time passes on the left boundary from $t_L$ to $\bar{t}_L$, the WDW patch starts growing in the green region and shrinks in the red region. At late times, the whole contribution of action growth comes from the green region that now lives between inner and outer horizons at $r=r_{\pm}$ of the charged black hole or between singularity at $r=r_s$ and future horizon of the neutral black hole. Therefore the total contribution of action comes from the bulk action in this region as well as the boundary actions on the space-like surfaces at $r_\pm$ or the singularity and future horizon.

\subsection{General quadratic curvature action} \label{GQC action}
Let us first introduce the bulk and boundary actions of GQC. The bulk action consists of the Einstein-Hilbert action together with a cosmological term. Moreover, we add all quadratic curvature terms as the sum of Ricci and Ricci scalar curvature squared terms and the well-known Gauss-Bonnet (GB) terms. To study charged black holes we also consider a $U(1)$ gauge field through a Maxwell term in our Lagrangian    
\be\label{LBulk}
I_{bulk}=-\frac{1}{2\k^2}\!\int_{\mathcal{M}}\!\!d^{d}x\sqrt{-g}\Big(R-2\L_0 +\mathit{a}_1 R^2 + \mathit{a}_2 R_{\m\n}^2 + \mathit{a}_3 (R^2 - 4 R_{\m\n}^2+R_{\m\n\r\s}^2)- F_{\m\nu}F^{\mu\nu}\Big),
\ee
where $\k^2=8\pi G$. 
To write the action for boundary surfaces, which make the variational principle of the gravitational field well-defined, instead of using the auxiliary field formalism (see e.g. \cite{Deruelle:2009zk}), we consider three types of Gibbons-Hawking (GH) terms corresponding to the Einstein-Hilbert, GB and Ricci square terms.
The GH term associated to the Einstein-Hilbert action is given by
\be\label{EHGH}
I_{GH}^{EH}=-\frac{1}{\k^2}\int_{\pa\mathcal{M}}\!\! d^{d-1}x \, \sqrt{-h} \, K\,,
\ee
where $h$ is the determinant of the induced metric on the boundary surfaces $\partial\mathcal{M}$ and $K$ is the trace of the extrinsic curvature. The extrinsic curvature is defined by 
\be 
K_{\mu\nu}=\frac12(\nabla_{\m} n_{\n}+\nabla_{\n} n_{\m})\,,
\ee
and $n_\m$ is a space-like unit vector, normal to the boundary.
Moreover, there is a generalized GH action corresponding to the GB gravity \cite{Myers:1987yn}
\be\label{GHGB}
I_{{GH}}^{GB}=\frac{2}{\k^2} \mathit{a}_3\int_{\pa\mathcal{M}}\!\!d^{d-1}x \, \sqrt{-h} \,\Bigl( 2 \Gg_{ab} {K}^{ab} +\tfrac{1}{3}(K^3 - 3 K K_{ab} K^{ab} + 2 K_{ab} K^{bc} K_{c}{}^{a} ) \Bigr),
\ee
where $\Gg_{ab}$ is the Einstein tensor constructed out of the induced metric. For Ricci squared terms of the action we use the following expression for GH term 
\be\label{RICGH}
I^{Ric^2}_{GH}=- \frac{1}{\k^2}\int_{\pa\mathcal{M}} d^{d-1} x \,\sqrt{-h}\, \frac{\partial L^{Ric^2}}{\partial R_{\a\m\b\n}} n_\mu n_\nu K_{\a\b}\,.
\ee
The total GH term therefore is the sum of these three parts
\be\label{stotGH}
I_{GH}=I^{EH}_{GH}+I_{{GH}}^{GB}+I^{Ric^2}_{GH}\,.
\ee

\subsection{Charged solution in d dimension} \label{Charged sol in d dim}
To find a charged solution with asymptotic AdS symmetries we use the following metric in $d$ dimensional space-time
\bea\label{anzats}
ds^2=-f_1(r) dt^2 + f_2^{-1}(r) dr^2 +r^2 d\Omega_{d-2}^2\,.
\eea
The field strength satisfies the Maxwell equation in $d$ dimension, i.e.  $\nabla_\m(\sqrt{-g}F^{\mu\nu})=0$
\be\label{ftr}
F_{tr}=-\sqrt{\frac12(d-2)(d-3)}\frac{q}{r^{d-2}}\sqrt{\frac{f_1(r)}{f_2(r)}}\,, 
\ee
where $q$ is the constant of integration, and is related to the electric charge $Q$ by
\be\label{charge}
q^2 =\frac{\k^2}{(d-2)\Omega_{d-2}}Q^2\,.
\ee
By variation of the Lagrangian (\ref{LBulk}) and inserting \eqref{anzats} and \eqref{ftr} for metric and field strength, we will find a couple of third and fourth order differential equations for $f_1(r)$ and $f_2(r)$. Finding an exact solution is a hard task to do, instead we find a solution which is linear in terms of the couplings of  theory i.e. $a_1, a_2$ and $a_3$. The solution is given by
\begin{subequations}
\begin{align}\label{f1sol}
f_1(r)&= 1-\frac{2\Lambda_0}{(d-1)(d-2)}\Big(1-\frac{C_0\Lambda_0}{(d-2)^2(d-1)}\Big)r^2 + q^2\Big(1-\frac{4\Lambda_0 C_q}{(d-2)(d-1)} \Big)\frac{1}{r^{2d - 6}}\nn \\ 
& - \frac{m}{r^{d-3}}+\frac{d-3}{d-2}\Big(C_1\frac{1}{r^{2 d - 4}}+2 q^2 m C_2\frac{1}{r^{3 d-7}} +q^4 \frac{C_3}{(3d-7)}\frac{1}{r^{4 d - 10}}\Big)\,,\\
\frac{f_2(r)}{f_1(r)}&=h(r)\,,\qquad h(r)=1-4\frac{d-3}{d-2} \big(a_1 (d-4) (2d-3)+a_2 (d^2-5d+5)\big) q^2 r^{4-2 d}\,,
\label{f2sol}
\end{align}
\end{subequations}
where all the above coefficients are as follow 
\begin{align}\label{coeff}
&C_q=a_3 (d-4) (d-3)+a_1 d(d-1)-a_2 (d(d-6)+7)\,, \nn \\ 
&C_0=2 (d-4) \big(a_3 (d-3)(d-2) + a_2 (d-1)+a_1
d(d-1)\big)\,, \nn \\ 
&C_1=\big(4 a_1 (d-4)-2 a_2 (d(d-6)+10)\big) q^2+a_3 (d-4) (d-2) m^2\,, \\\ 
&C_2=-a_1 (d-4) (d-1)+a_2-a_3 (d-4) (d-2)\,, \nn \\ 
&C_3= (d-4)( a_3(d-2) (3d-7)+a_1(11d^2-45d+44))+a_2 (4d^3-33d^2+83d-64)\,.\nn
\end{align}
To prevent the divergence in the gauge field strength \eqref{ftr}, the function $h(r)$ in relation \eqref{f2sol}  should not be equal to zero. This requires considering the following condition between couplings of the theory
\be
a_1 (d-4) (2d-3)+a_2 (d^2-5d+5)\leq 0\,.
\ee

For the neutral solution as we send $q\rightarrow 0$, both functions in \eqref{f1sol} and \eqref{f2sol} are equal or $h(r)=1$ at this level of perturbation, but it can be shown that, $f_1(r)$ and $f_2(r)$ are separated when we consider next orders of perturbation into account.

\subsection{Action growth rate in WDW patch} \label{Acgro in WDW patch}
The growth rate of bulk action in the WDW patch at late-time approximation can be computed by inserting the solution \eqref{f1sol} and 
\eqref{f2sol} into bulk action \eqref{LBulk}. For more details of computations see relations \eqref{use1}$-$\eqref{use3} in Appendix  \ref{appendix A}  
\bea\label{grbulk}
\!\!\!\!\!\!\!\!\frac{dI_{bulk}}{dt}=\frac{\Omega_{d-2}}{2\k^2}\int_{r_-}^{r_+} \!\!\! dr \,\frac{2 r^{-2-3d}}{(d-2)^3(d-1)} \Big(\a_1 r^{2d+4} +\a_2 r^{2d+2}+\a_3 r^{d+5}+\a_4 r^{4d}+\a_5 r^{8}\Big)\,,
\eea
where $\a_1,...,\a_5$ are given in equation (\ref{alphacoef}) in Appendix  \ref{appendix B}. In calculation of action growth in this section we consider $r_{\pm}$ as outer/inner horizons and we suppose that there is no singularity in between them.

At late-time approximation, the growth rate of Gibbons-Hawking surface terms of WDW patch is obtained by computing the value of \eqref{stotGH}. To do this we insert the solution \eqref{f1sol} and \eqref{f2sol} into the equations \eqref{use4} and \eqref{use5} and we find
\begin{align}\label{grbound}
\frac{dI_{GH}}{dt}&=-\frac{\Omega_{d-2}}{\k^2}\frac{1}{6(d-2)^3(d-1)(3d-7)}\Big(\b_1 r^{-3d+7}+\b_2 r^{d-1}+\b_3 r^{d-3}+\b_4 r^{d-5}\nn \\
&+\b_5 +\b_6 r^{-2}+\b_7 r^{-d+3}+\b_8 r^{-d+1}+ \b_9 r^{-2d+4}\Big)\Bigg|_{r_-}^{r_+},
\end{align}
where $\b_1,...,\b_9$ are given in equation (\ref{betacoef}).
Finally, we perform the integration of \eqref{grbulk} and add it to the relation \eqref{grbound} which leads to the total action growth rate. The final result can be simplified  by using the following steps:

1. Because we considered $r_+$ as the outer horizon, we can solve $f_1(r_+)=0$ to find a relation for $\L_0$.

2. Insert the value of $\L_0$ from the first step into $f_1(r_-)=0$ to find a relation for $m$ in terms of $r_+$ and $r_-$.

3. Insert the relations of $\L_0$ and $m$ in the first and second steps into the total action growth rate. 
The final result is
\begin{align}\label{dstotdt}
\frac{dI_{tot}}{dt}&=-\frac{\Omega_{d-2}}{\k^2}q^2\Big((d-2)\big(\frac{1}{r_-^{d-3}}-\frac{1}{r_+^{d-3}}\big)\nn \\
&+\frac{2(d-3)^2q^2}{3d-7}\big(a_1 (2d-3)(d-4) +a_2 (5-5d+d^2)\big)  \big(\frac{1}{r_-^{3d-7}}-\frac{1}{r_+^{3d-7}}\big)\Big) \,.
\end{align}
Using the field strength relation \eqref{ftr} and the value of electric charge in \eqref{charge}, one can write the above expression as follow
\be\label{dsdtc}
\frac{dI_{tot}}{dt}=(M-\mu_+ Q)-(M-\mu_- Q)\,,
\ee
where we have supposed $M$ as the mass of black hole and
\be
\mu_\pm=-\frac{Q}{r_{\pm}^{d-3}}-\frac{\k^2}{\Omega_{d-2}}\frac{2(d-3)^2}{(d-2)^2(3d-7)}\Big(a_1 (2d-3)(d-4)+a_2 (5-5 d+d^2)\Big)\frac{Q^3}{r_{\pm}^{3d-7}}\,,
\ee 
are the values of chemical potential  on $r_\pm$ horizons.

The result of equation \eqref{dsdtc} shows that the proposal introduced in \cite{Cai:2016xho} is correct for general quadratic curvature theory.  Using the Noether charge formalism of Iyer and Wald, \cite{Jiang:2018pfk} finds this modified version of the Lloyd's bound in multiple Killing horizons black hole for a higher curvature theory of gravity.

\subsection{Neutral black hole in d dimension}\label{Neutral BH in d dim}
To find the action growth rate in the neutral case we need to know the singular points of the geometry.
To do this we first review the solutions of equations of motion for some special cases:

$\bullet\,\,\, a_3=0$: In this case the exact solution of equations of motion is given by the known Schwarzschild AdS black hole, for example see \cite{Ghodsi:2017iee}
\begin{subequations}
\begin{align}
&f_1(r)=f_2(r)=1+\frac{r^2}{L^2}-\frac{m}{r^{d-3}}\,, \\
& -(d-2) (d-1) \frac{1}{L^2}+(d-4) (d-1)^2 (a_2+a_1 d) \frac{1}{L^4}-2 \Lambda_0=0\,.
\end{align}
\end{subequations}
This solution has a singularity at $r_s=0$.

$\bullet\,\,\, a_1=a_2=0$: This case corresponds to the Gauss-Bonnet gravity. The exact solution is given by \cite{Cai:2016xho}
\begin{subequations}
\begin{align}\label{GBsol}
f_1(r)&=f_2(r)=1+\frac{r^2}{\l} \Big(1-\sqrt{1+2\l\big(\frac{2 \Lambda_0}{(d-1)(d-2)}+\frac{m}{r^{d-1}}}\big)\Big)\,,\\
\l&=2 a_3 (d-4) (d-3) \,.
\end{align}
\end{subequations}
This solution has two singularities. The first one, is where the metric is divergent and the other one, is the point where the metric or scalar curvature terms become imaginary
\be\label{sings}
r_s=0\,,\qquad r_s= \Big(\!\!-\frac{ 2\l (d-2) (d-1)}{(d-1)(d-2)+4\l \Lambda_0}m\Big)^{\frac{1}{d-1}}\,.
\ee
The above analysis for special cases helps us to find the singularity structure of the solution for general case. To make the analysis easier, it would be better to find a solution which reduces to solution \eqref{f1sol} when $q\rightarrow 0$ and reduces to \eqref{GBsol} when $a_1=a_2=0$. 

In fact, we may assume that there is a  general solution, which can be written as
$f(r)=1+r^2 X(r)\Big(1-\sqrt{1+c+\frac{m}{r^{d-1}}}\Big)$
and solve equations of motion perturbativly to find $c, m$ and $X(r)$.
To first order of perturbation both $f_1(r)$ and $f_2(r)$ functions are given by
\bea\label{exsol}
&&\!\!\!\!\!\!\!\!\! f(r)=1+\frac{1-\frac{2 (d-4) (a_2+d a_1) \Lambda_0}{(d-2)^2} }{\l} r^2\Big(1-\sqrt{1+2\l \big(\frac{2 \Lambda_0}{(d-1)(d-2)}+\frac{m}{r^{d-1}}}\big)\Big)\,.
\eea
We can see that \eqref{exsol} has the same the singularity structure as \eqref{GBsol} at least at this order of perturbation.
By an expansion around small couplings, \eqref{exsol} would be equal to
\begin{subequations}
\begin{align}\label{fsoln}
f(r)&= 1+\frac{r^2}{L^2} - \frac{m}{r^{d-3}}+ \frac{\l}{2}\frac{m^2}{r^{2 d - 4}}\,,\\
\frac{1}{L^2}&=-\frac{2\Lambda_0}{(d-1)(d-2)}\Big(1-\frac{(d-2)\l +2 (d-1)(d-4)( a_2 +d a_1) }{(d-2)^2 (d-1)}\Lambda_0\Big)\,,
\end{align}
\end{subequations}
where in this expansion $m=m_0+\d m$ ($m_0$ is a constant of integration) and 
\be
\d m=\frac{-2}{(d-2)^2 (d-1)}\Big((d-2)\l +(d-1) (d-4)(a_2+d a_1)\Big) \Lambda_0 m_0\,.
\ee
\subsection{Action growth rate in WDW patch}\label{Neutral Acgro rate}
To find the growth rate of actions for neutral Schwarzschild AdS black hole we carry out the following steps:

1. First, we add equations \eqref{grbulk} and \eqref{grbound} after taking the limit of $q \rightarrow 0$, with a distinction that the upper limit is $r=r_h$, the location of the future horizon and the lower limit is $r=r_s$, the location of the singularity
\begin{align}\label{nrate}
\frac{dI^{q=0}_{tot}}{dt}&=\frac{\Omega_{d-2}}{6\k^2}
\Big(3(d-1)m
-(d-2) m^2 r^{1-d} \lambda+2 m \big(\frac{(d-2)\lambda}{r^2}+\frac{6 (d-1) (a_2+a_1 d)\Lambda_0}{d-2}\nn \\&
+\frac{\lambda \Lambda_0}{d-4}\big)+\frac{2r^{d-5}}{(d-1)^2(d-2)^2}\big(2 (d-2)^2 (d-1) r^2 (3 r^2+\!\lambda-6 (d-1) (a_2+a_1 d)) \Lambda_0\nn \\
&-(d-2)^3 (d-1)^2 (3 r^2+2 \lambda)+2 r^4 (6 (d-1) d (a_2+a_1 d)-(d-2) \lambda) \Lambda_0^2\big)
\Big)\Big|_{r_s}^{r_h}.
\end{align}

2. The horizon is given by solving the equation
\be\label{lhor}
1+\frac{r_h^2}{L^2} - \frac{m}{r_h^{d-3}}+ \frac{\l}{2}\frac{m^2}{r_h^{2 d - 4}}=0\,.
\ee
We use this equation as a constraint between $r_h$ and other parameters.
Suppose that the location of horizon from this equation can be found perturbativly by putting $r_h=r_0+\d r$ into \eqref{lhor}. In this way, we can find the following relations
\begin{subequations}
\begin{align}
\d r&=\frac{1}{-(d-3) m_0 r_0^{2-d}+\frac{4 r_0 \Lambda_0}{(d-1)(d-2)}}\Big(\frac12 \l m_0^2 r_0^{4-2 d}-\d m r_0^{3-d}\nn \\ 
&+\frac{ 2\big(\l (d-2)+2 (d-1)(d-4)(a_2 +a_1 d)\big) r_0^2 \Lambda_0^2}{(d-2)^3 (d-1)^2}\Big)\,,\\
1&-m_0 r_0^{3-d}-\frac{2 r_0^2 \Lambda_0}{(d-1)(d-2)}=0\,.
\end{align} 
\end{subequations}

3. Using the above equations we can simplify the growth rate  \eqref{nrate} at $r=r_h$. For simplicity from now on we drop the constant (independent of $r_h$) terms in \eqref{nrate} (note that they will cancel when we compute the total growth), then we have
\begin{align}
\frac{dI^{q=0}_{tot}}{dt}\Big|^{r_h}&=-\frac{m\Omega_{d-2}}{\k^2}\Big((d-2)+\frac{2(d-3)\Lambda_0}{(d-1)(d-2)}(a_3 (d-5) (d-2)+(d-1)(a_2 +a_1d)) \Big)\,.
\end{align}

4. In general the action growth rate at $r=r_s$ diverges when $r_s\rightarrow 0$. Nevertheless, there are special cases that its value is finite:

$\bullet$ For $a_3=0$, as we already mentioned, the singularity is located at $r_s=0$. At this value the equation \eqref{nrate} will be equal to 
\be
\frac{dI^{q=0}_{tot}}{dt}\Big|_{r_s=0}=\frac{2m\Omega_{d-2}}{\k^2}\frac{d-1}{d-2}(a_2+a_1 d)  \L_0\,,
\ee
therefore the total action growth rate would be equal to
\be\label{agra30}
\frac{dI^{q=0}_{tot}}{dt}=-\frac{m\Omega_{d-2}}{\k^2}\Big((d-2)+4(a_2+a_1 d)\L_0\Big)=2M\,.
\ee
The last equality is written in terms of the mass of black hole for the Ricci square theory of gravity. For example see \cite{Ghodsi:2017iee} for more details of computation of this mass\footnote{To translate computations of \cite{Ghodsi:2017iee} to ours, we need to change $a_1 \leftrightarrow a_2$, $\k^2\rightarrow -\k^2$ and replace $l=1$, $r_0^{d-1}=m$, $V_{d-2}=\Omega_{d-2}$, $\bar{\s}=1+2(d-1)\Lambda(a_2+a_1 d)$ and $\Lambda_0=\frac12(d-1)\Lambda(d-2+(d-4)(d-1)\Lambda(a_2+da_1))$ in mass equation (5.13) i.e. $M=\frac{(d-2)V_{d-2}}{8\pi G_d l}(\frac{r_0}{l})^{d-1}\bar{\s}$.}.
Note that \eqref{agra30} is a general result and for example as a special case it gives the result of $d$ dimensional critical gravity.

$\bullet$ For $d=4$ the Gauss-Bonnet part of the theory is a topological term. The growth rate in this case is equal to
\begin{align}
\frac{dI^{q=0}_{tot}}{dt}&=-\frac{4\pi^2 m}{\k^2}\Big((2+4a_1+a_2-\frac23 a_3)-(-12a_1-3a_2-\frac23 a_3)\Big)\nn\\
&=-\frac{8\pi^2 m}{\k^2}\Big(1+2(a_2+4a_1)\L_0\Big)=2M\,.
\end{align}

5. For a more general case when $a_3< 0$,  there is a second singularity at $r_s\neq 0$. This can be seen from equation \eqref{sings} for $m>0$ and $\L_0<0$. Although this new point of singularity makes the equation  \eqref{nrate} finite, but the final result leads to a wrong answer for  late time complexity 
\be
\frac{dI^{q=0}_{tot}}{dt}=-\frac{(d-2)\Omega_{d-2}}{\k^2}\Big(\frac{13}{12}m+\frac{7}{3\times 4^{d-1}} \big(-2(d-3)(d-4)a_3 m\big)^{\frac{d-3}{d-1}}\Big)\,.
\ee
A similar behavior also has been reported for the $d=5$ dimensional Gauss-Bonnet gravity in \cite{Cai:2016xho}.
\subsection{More on the Second singularity and complexity}
In previous subsections we studied the GQC black hole's solution perturbativly. We assumed two outer and inner horizons and singularity was located behind the inner one. As we investigated (perturbativly) in neutral black holes, it is possible to have a second singularity which  leads to a wrong answer for late time complexity. In this subsection we will try to study $d$-dimensional charged black hole solution in Gauss-Bonnet gravity which can be found analytically and contains a second singularity as well, (see \cite{Cai:2016xho} for $d=5$). 
The metric \eqref{anzats} after solving the equations of motion is given by $f_1(r)=f_2(r)=f(r)$
\bea\label{exqsol}
&&\!\!\!\!\!\!\!\!\! f(r)=1+\frac{r^2}{\l}\Big(1-\sqrt{1+2\l \Big(\frac{2 \Lambda_0}{(d-1)(d-2)}+\frac{m}{r^{d-1}}-\frac{q^2}{r^{2d-4}}}\Big)\Big)\,.
\eea
Then the location of horizons and the second singularity (interval) are given by the following algebraic equations
\be 
f(r_h)=0\,,\qquad 1+2\l \Big(\frac{2 \Lambda_0}{(d-1)(d-2)}+\frac{m}{r_s^{d-1}}-\frac{q^2}{r_s^{2d-4}}\Big)=0\,.
\ee
In general, finding the analytic solutions for above equations is a hard task to do, but a simple numerical investigation shows the possible situations. In figure \eqref{Fig3} we have sketched the various possible solutions for $f(r)$.

The asymptotic AdS nature of solution restricts us to consider $\L_0<0$. Moreover, $m$ is proportional to the mass of black hole so $m>0$. The Gauss-Bonnet coupling could be either negative or positive. In figure \eqref{Fig3} we have fixed $\L_0=-0.01$, $m=1$ and $a_3=\mp 0.02$ in the left/right diagrams, in $d=5$. For various values of the charge $q$, we have the following behaviors:
\begin{figure}[!hbt]
	\centering
	\includegraphics[scale=0.51]{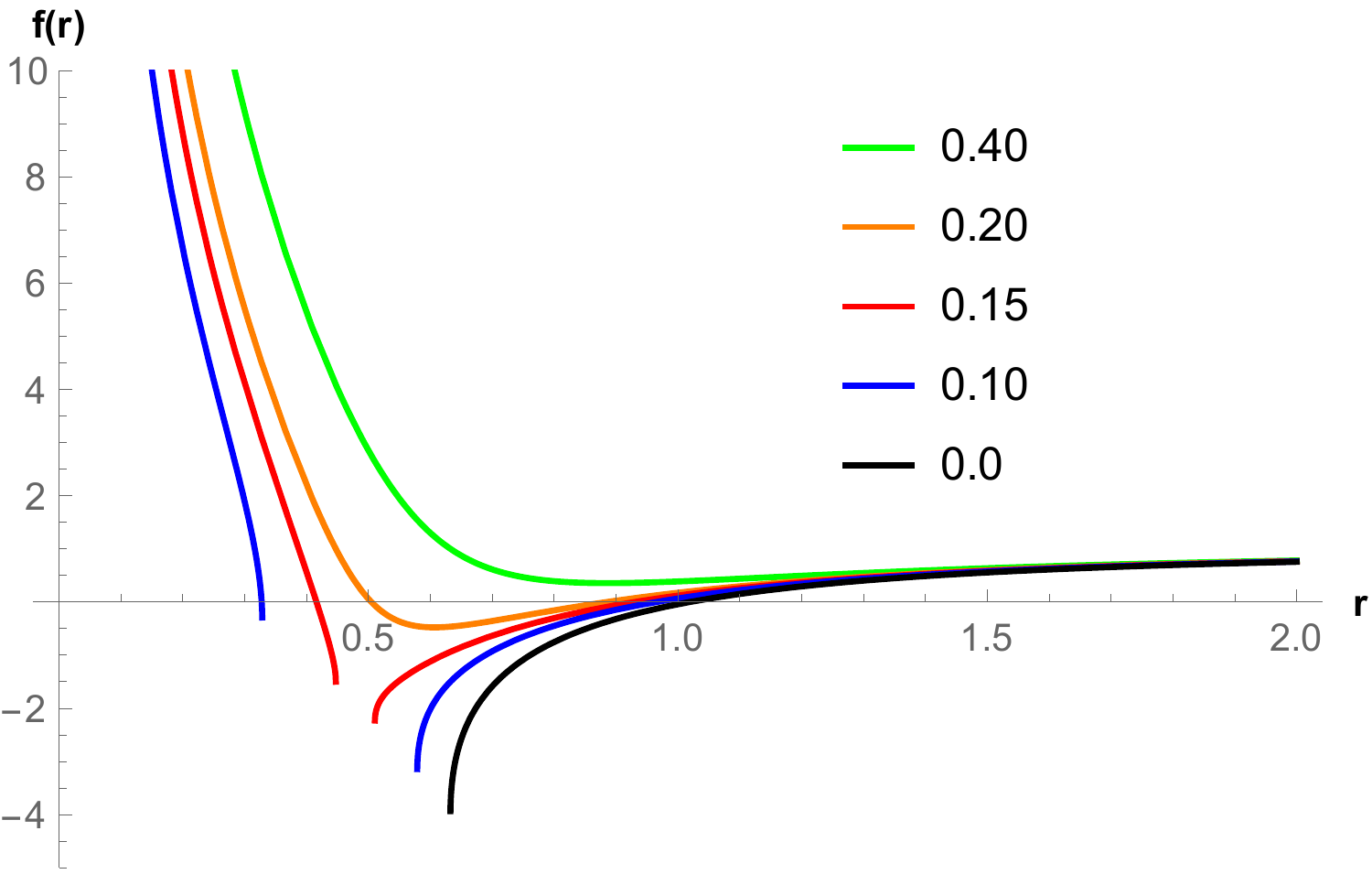}
	\includegraphics[scale=0.51]{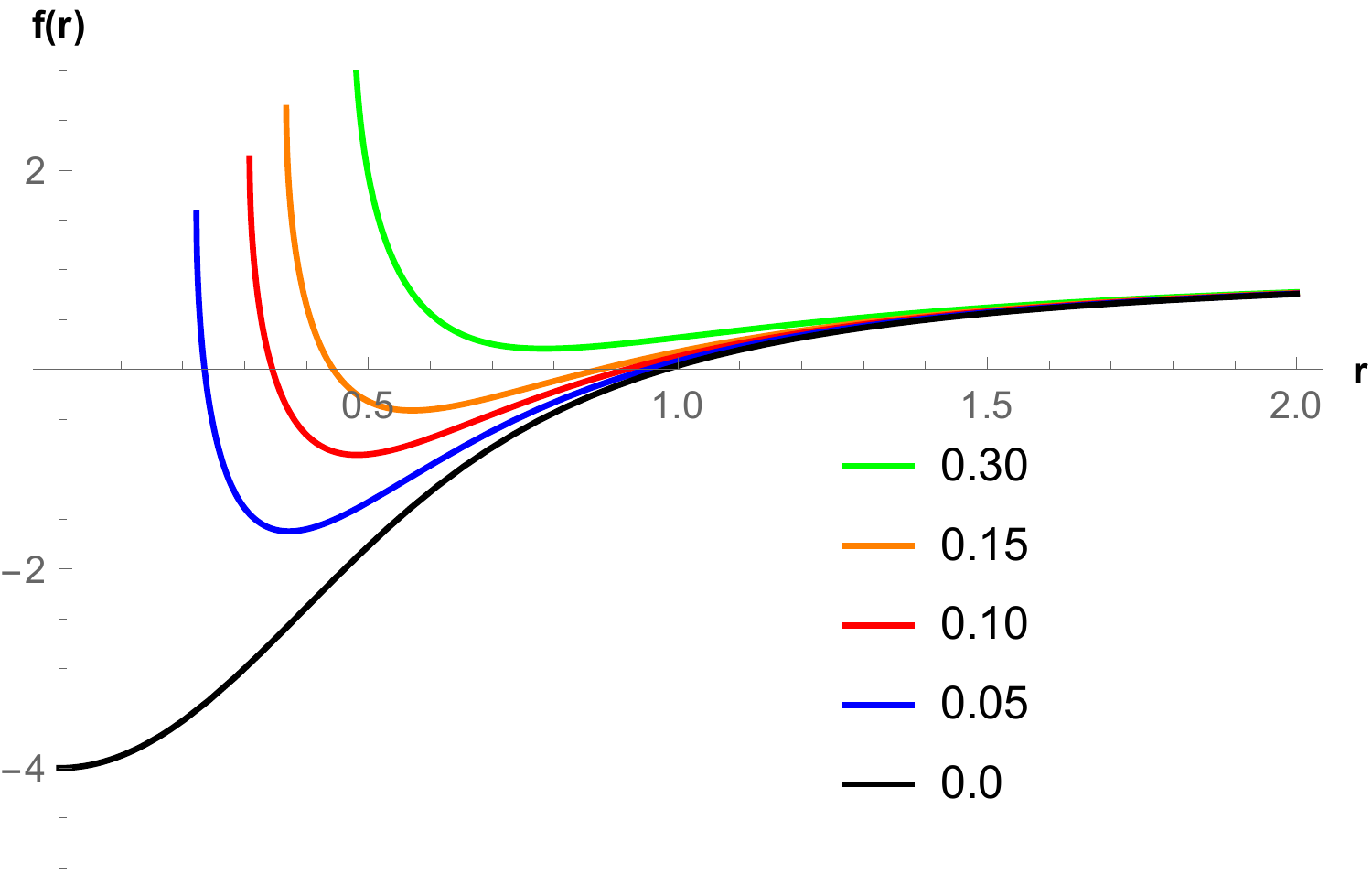}	
	\caption{\footnotesize{Horizon and singularity structure of charged Gauss-Bonnet black holes.}}
	\label{Fig3}
\end{figure}

The left diagram:

1. There is a naked singularity at $r_s=0$, green curve.

2. There are two horizons and a singularity at $r_s=0$, orange curve.

3. A second singularity at $r_s\neq 0 $ exists between two horizons, red and blue curves.

4. The second singularity at $r_s\neq 0 $ exists behind the horizons, black curve (neutral solution).

The right diagram:

1. A singularity at $r_s\neq 0 $ exists behind the inner horizon, green to blue curves.

2. A solution with a horizon and no singularity, black curve (neutral solution).

By performing the same steps as we did in the previous sections we find the following relation for total action growth:

When the second singularity is behind the inner horizon the situation is similar to the perturbed solution which we found already 
\be 
\frac{dI_{tot}}{dt}=\frac{\Omega_{d-2}}{\k^2}q^2(d-2)\big(\frac{1}{r_+^{d-3}}-\frac{1}{r_-^{d-3}}\big)=
(M-\mu_{+} Q)-(M-\mu_{-} Q)\,.
\ee
On the other hand when the second singularity is between the two horizons
\be 
\frac{dI_{tot}}{dt}=\frac{\Omega_{d-2}}{\k^2}\Big(q^2(\frac{1}{r_h^{d-3}}-\frac{1}{r_s^{d-3}})+\frac{2(d-2)r_s^{d-5}}{3\l}(r_s+\l)^2\Big)\,.
\ee
As we see, although the first term can be written as $(M-\mu_h Q)-(M-\mu_s Q)$, but there is an extra term which leads us to  another value for the Lloyd's bound.
\section{WDW action for global AdS} \label{universal terms}
In this section, we are going to compute the universal terms that appear in the divergent part of $C_A$ complexity in the GQC theory of gravity. All steps that we follow here, have been presented already in reference \cite{Cano:2018ckq}. We will show how with some simple modifications we can find the universal coefficients of $C_A$. 

Paper \cite{Cano:2018ckq} begins with Lovelock theory with the following bulk and space/time-like boundary actions 
\be\label{lovelock}
I=\sum_n \l_n \left(\int_M d^d x \sqrt{-g}\, \mathcal{X}_{2n}+\int_{\cup_k B_k} \!\!\!\! d^{d-1} x\, d\Sigma\, \mathcal{Q}_n\right)\,,
\ee
where $\mathcal{X}_{2n}$ is the Euler density and $\mathcal{Q}_n$ is the generalized GH boundary term. The boundary terms make the variational principle well defined. For example for $n=1,2$, the relations for $\mathcal{Q}_1$ and $\mathcal{Q}_2$ are given in equations \eqref{EHGH} and \eqref{GHGB}.

After that, \cite{Cano:2018ckq} computes the contribution from space/time-like joints by employing the Hayward smoothing method. These joints are co-dimension two surfaces, which are made from the intersection of boundary surfaces. In this way \cite{Cano:2018ckq} prove that the joint terms can be computed from the Lovelock boundary term in \eqref{lovelock} and are given by
\begin{equation}\label{joint1}
I_{joint}=\sum_n \l_n\int_{C}d\sigma 2n \eta \hat{\mathcal{X}}_{2(n-1)}\,,
\end{equation}
where $\eta=\pm cosh^{-1}|n_1.n_2|$, and $n_{1,2}$ are normal one-forms to each boundary that their intersection makes the joint $C$. In above equation $\hat{\mathcal{X}}_{2(n-1)}$ is the Euler density constructed from  the induced metric on the joint. 

\begin{figure}[!hbt]
	\centering
	\includegraphics[scale=1]{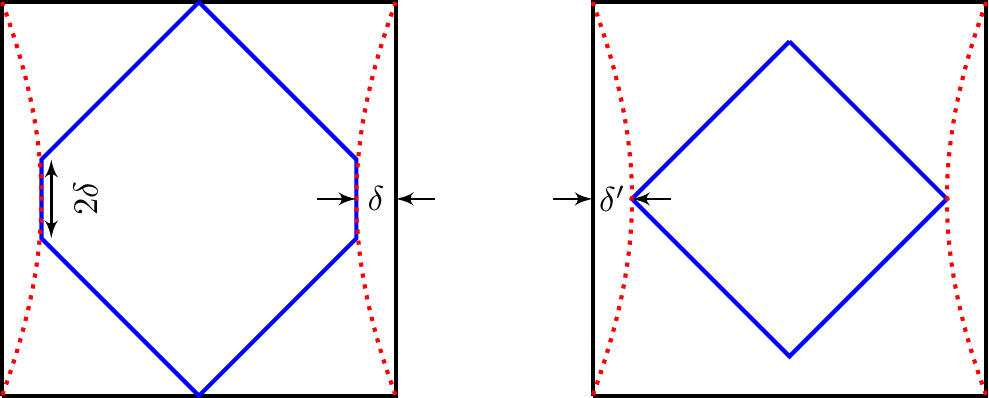}	
	\caption{\footnotesize{Two ways for regularizing the WDW patch. }}
	\label{Fig2}
\end{figure}

To find the universal terms, one needs to compute the gravitational action on a regularized WDW patch. This patch contains a cut-off distance $\delta$ from the boundary (see figure \eqref{Fig2})

By choosing this patch, \cite{Cano:2018ckq} argues that despite of surface terms for null boundaries, the universal terms  are originating either from the joint terms \eqref{joint1} or the bulk action. The final form of joint term is given by
\begin{equation}\label{joint2}
I_{joint}=-\frac{1}{\k^2}\int_{C}d\sigma a \Bigg[1+\!\!\sum_{n=2}^{[ (d-1)/2]} n \lambda_n \hat{\mathcal{X}}_{2(n-1)}\Bigg]\,,
\end{equation}
where $a=\pm\log|k_1.k_2/2|$ for joints between two null boundaries and  $a=\pm\log|n.k_1|$ for joints between a time-like and a null boundary.

Now consider the GQC action \eqref{LBulk} without Yang-Mills term. Although the bulk action cannot be written as a sum of Euler terms but as it was shown in \cite{Ghodsi:2019xrx} or from computing equation \eqref{RICGH} for a Global AdS solution, the total GH boundary term \eqref{stotGH} can be written as
\begin{align}\label{effGH}
I_{GH}&=\l_{eff} I^{EH}_{GH}+ I_{{GH}}^{GB}\,,\qquad
\l_{eff}=1+4\Lambda_0 \frac{d a_1+a_2}{d-2}\,.
\end{align}
Therefore the joint term \eqref{joint2} for GQC theory simplifies to
\begin{equation}\label{GQCjoint}
I^{GQC}_{joint}=-\frac{1}{\k^2}\int_{C}d\sigma a \Big[\l_{eff}+ 2 a_3 \hat{\mathcal{R}}\Big]\,,
\end{equation}
where $\hat{\mathcal{R}}$ is the scalar curvature on the joint that is constructed from the induced metric. 

Following \cite{Carmi:2016wjl}, we use the regulated graph depicted in figure \eqref{Fig2}, i.e. we change the WDW patch by an inward shift on the right and left edges. 
To compute the structure of divergences, we begin with the following metric which is asymptotically $AdS_d$ space-time with radius $\tilde{L}$ in GQC theory
\begin{subequations}
\begin{align}
ds^2&=-f(r) dt^2 + f^{-1}(r) dr^2 +r^2 d\Omega_{d-2}^2\,,\qquad 
f(r)\Big|_{r \to \infty} \approx 1+\frac{r^2}{\tilde{L}^2}\,, \\
\frac{1}{\tilde{L}^2}&=\frac{-2\L_0}{(d\!-\!1)(d\!-\!2)}\Big(1-\frac{2\L_0(d\!-\!4)}{(d\!-\!1)(d\!-\!2)^2}\big((d\!-\!1)( d a_1 + a_2) + a_3 (d\!-\!2)(d\!-\!3)\big)\Big)\,.
\end{align}
\end{subequations}
By two successive proper coordinate transformation i.e. $z=\frac{\tilde L}{r}$ and $t=\tau \tilde L$, and then  $z=\frac{2\tilde L \,\cos\theta}{1+\sin\theta}$
we find the following metric
\begin{equation}
ds^2 = \frac{\tilde L^2}{\cos^2\!\theta}\left(-d\tau^2+d\theta^2+\sin^2\!\theta\,d\Omega_{d-2}^2\right)\,.
\end{equation}
In this metric, the null boundaries of the WDW patch (the right diagram of figure \eqref{Fig2}) are given by
\begin{subequations}
\begin{align}
S^+ :&\quad \theta= \frac{\pi}{2}-\tau -\d' \qquad  {\rm for} \qquad \frac{\pi}2-\d' \ge\tau\ge 0\,, 
\\
S^- :&\quad \theta= \frac{\pi}{2}+\tau -\d' \qquad {\rm for}\quad  -\frac{\pi}2 +\d '\le\tau\le 0\,. 
\end{align}
\end{subequations}
Moreover the unit normal vectors to $S^+$ and $S^-$ null surfaces are
\be
k_1=\a_1\,\tilde L\,(d\theta+d\tau)\,,\qquad 
k_2=\a_2\,\tilde L\,(d\theta-d\tau)\,,\qquad 
\ee
where $\a_{1,2}$ are normalization constants. By another change of coordinate, $\theta'=\frac\pi2-\theta$ the location of joint is at $(\t,\theta')=(0,\d')$. In this way the induced metric on the joint and scalar curvature tensor would be
\begin{equation}
d\hat{s}^2=\tilde L^2 \cot^2\!\delta' \,d\Omega_{d-2}^2\,,
\qquad
\hat{\mathcal{R}}=(d-2)(d-3)\frac{\tan^2\!\d'}{\tilde L^2}\,.
\end{equation}
Also for a joint between two null boundaries in this coordinate
\bea
a= -\log|\frac{k_1.k_2}{2}|=-\log (\alpha_1\alpha_2\sin^2\!\d') \,. 
\eea
By substituting the above values into the joint action \eqref{GQCjoint}, finally, we obtain
\bea\label{joint4}
I_{joint}=
\frac{2\tilde L^{d-2}\Omega_{d-2}}{\k^2 \tan^{d-2}\!\delta'} \log\big(\sqrt{\alpha_1\alpha_2}\sin\d'\big) \left[\l_{eff}+2 a_3\frac {(d-3)(d-2)}{\tilde L^2} \tan^2\!\d'\right]\,.
\eea 
We can write the above expression in terms of original cut-off by using $\d=\frac{2\tilde{L}\sin\d'}{1+\cos\d'}$. After expansion around $\d=0$ we find
\begin{align}
&I_{joint}=-\frac{2\tilde L^{d-2}\Omega_{d-2}}{\k^2 }\Big( \l_{eff}N_1 +2 a_3\frac {(d-3)(d-2)}{\tilde L^2} N_2\Big)\log\big(\frac{\tilde{L}}{\sqrt{\alpha_1\alpha_2}\d}\big)+\cdots\,, \\
&N_1=\sum_{n=0}\frac{(-\frac14)^n\G(d-1)}{\G(d-1-n)\G(n+1)}(\frac{\tilde{L}}{\d})^{d-2-2n}\,,\quad N_2=\sum_{n=0}\frac{(-\frac14)^n\G(d-3)}{\G(d-3-n)\G(n+1)}(\frac{\tilde{L}}{\d})^{d-4-2n}\,,\nn
\end{align}
where dots represent the power expansion terms.
Therefore the universal term for GQC from the joint term for even $d$ dimensional space-time is given by
\begin{subequations}
\begin{align}
&C^{univ}_{joint}=(-1)^{\frac{d-2}{2}}b_d \log\big(\frac{\tilde{L}}{\sqrt{\alpha_1\alpha_2}\d}\big)\,, \\
&b_{d}=-\frac{2\tilde L^{d-2}\Omega_{d-2}}{\pi\hbar\k^2 }\frac{\G(d-1)}{2^{d-2}\G(\frac{d}{2})^2}\Big(\l_{eff} -\frac {(d-2)^2 a_3}{2\tilde L^2} \Big)\,.\label{UTJ}
\end{align}
\end{subequations}
We can also find the universal terms, which obtain from the  computation of the bulk action on WDW patch. In $d$ dimensional space-time these terms are computed in \cite{Cano:2018ckq}
\begin{align}
C^{univ}_{bulk}=\left\{
\begin{array}{ll}
(-1)^{\frac{d-3}{2}} \frac{2}{\hbar} a^{*}_d & \qquad {\rm for \  odd \ d}\,, \\ \\
(-1)^{\frac{d-4}{2}} \frac{4}{\pi\hbar} a^{*}_d \log\frac{\tilde{L}}{\d} & \qquad {\rm for \  even \ d}\,,
\end{array}
\right.
\end{align}
where $a^{*}_d$ is the $a$-anomaly, and for GQC theory it is equal to (see \cite{Ghodsi:2019xrx} for more discussions)
\begin{align}
\label{a*GQC}
a^*_{d} &= \frac{\pi^{\frac{d-1}{2}} \td{L}^{d}}{(d-1)\, \Gamma(\frac{d-1}{2})} \mathcal{L}_{bulk}\Big|_{AdS}\nn \\
& = \frac{\pi^{\frac{d-1}{2}} \td{L}^{d-4}}{\k^2 \Gamma(\frac{d-1}{2})} \Bigl( \td{L}^2 - 2\big( (d-1)( d a_1 + a_2) + a_3 \,(d-2)\,(d-3)\big)\Bigr)\,.
\end{align}

\section{Conclusion and Discussion} \label{Summary}
In this paper we use the CA conjecture \eqref{CAC} for studying the holographic complexity. In the bulk space-time, we assume a general quadratic action that includes both Riemann and Ricci curvature tensors up to the quadratic terms \eqref{LBulk}. This form of action allows us to generalize the ideas around the holographic complexity when one takes into account the higher curvature theories of gravity in $d$ dimensional space-time. 

In section \ref{C-dot in GQC} we examine the proposal of growth bound on complexity for two types of charged and neutral black holes of GQC that are asymptotically AdS space-time. In subsection \ref{Charged sol in d dim} we first find a $U(1)$ charged black hole solution and then in subsection \ref{Acgro in WDW patch} we obtain the action growth rate \eqref{dstotdt} by computing the bulk and boundary actions on WDW patch at late-time approximation. Our final result \eqref{dsdtc} confirms the proposal in \cite{Cai:2016xho}, i.e. we can write the total action growth as a difference between the value of $M-\mu Q$ on the outer and inner horizons of the black hole. This result has been already reported for various theories of gravity \cite{Jiang:2018pfk},
 and here we observe that this bound is preserved for general quadratic curvature theory of gravity.

Despite this, the case of the neutral black hole is more challenging. The reason is the existence of the singularity as one of the surface boundaries of the WDW patch. In subsection \ref{Neutral BH in d dim} we learn how we can guess and compute the singularity structure of geometry in GQC theory from a particular class of solutions . Using this, in subsection \ref{Neutral Acgro rate} we compute the action growth rate. We show that for special cases such as $a_3=0$ or $d=4$ where the singularity is located at $r_s=0$, the singularity does not produce any divergences and the Lloyd's bound saturates at $2M$. 

The above results hold when the singularity point is located behind the horizon(s).
A numerical survey shows that, for certain regions of the parameters of the theory a second singularity appears at $r_s\neq 0$. For neutral black holes this changes the expected proposal of  \cite{Cai:2016xho}.
Moreover for charged black holes when this singularity is behind the inner horizon, the action growth rate saturates the Lloyd's bound. But when it is located between the two horizons then it leads to a result other than the expected Lloyd's bound.

It is worth reminding that, even though higher order curvature theories of gravity may contain the ghost modes, as indicated in \cite{Yang:2016awy} it can be shown that the strong energy condition is a sufficient condition to ensure the bound inequality, therefore in this paper, we suppose this energy condition everywhere, (for more details on energy conditions of the GQC see \cite{Ghodsi:2019xrx}).

In section \ref{universal terms} we look into another interesting subject in the context of holographic complexity and compute the universal terms that appear in the divergent part of $C_A$ for the GQC theory of gravity. Usually, there are two types of these universal terms, one from the joint terms of regularized WDW patch and one from the bulk action. Using a simple trick, by introducing an effective GH term \eqref{effGH}, we find the joint terms from the techniques in paper  \cite{Cano:2018ckq}. In that paper, the joint terms were calculated for Lovelock theory. Although the GQC cannot be written in terms of the Lovelock theory, its GH terms as we mentioned, are compatible with those in \cite{Cano:2018ckq} technique. In this way, we find the universal terms \eqref{UTJ} and \eqref{a*GQC}.   
\section*{Acknowledgment}
We would like to thank M. Siahvoshan for useful discussion. This work is supported by Ferdowsi University of Mashhad under the grant 2/52193.
\appendix
\section{Useful relations}\label{appendix A}
Using the anzats in \eqref{anzats} we find the following useful expressions for computing the bulk Lagrangian \eqref{LBulk}
\begin{subequations}
\begin{align}\label{use1}
R&=-\frac{r^{-2}}{2 f_1^{2}}\Big(2 (d-2) f_1^2 \big((d-3) (f_2\!-\!1)+r f'_2\big)-r^2f_2 {f'_1}^2\nn \\
&+rf_1  \big(r f'_1 f'_2+2  \big((d-2) f'_1+r f''_1\big)f_2\big)\Big)\,, \\\label{use2}
R_{\m\nu}R^{\m\n}&=\frac{r^{-4}}{8 f_1^{4}}\Big(2 (d-2) f_1^4 \big(4 (d-3)^2 (f_2-1)^2+4 (d-3) (f_2-1) r f'_2+(d-1) r^2 {f'_2}^2\big)\nn \\
&+r^4f_2^2  {f'_1}^4\!-\!2 r^3 \big(f'_1 \big((d\!-\!2) f_2+r f'_2\big)+2r f_2  f''_1\big) f_1 f_2{f'_1}^2\!+\!2 (d\!-\!2)  r f_1^3\big(2r^2  f_2 f'_2 f''_1\nn \\
&+f'_1 \big(4 (d\!-\!3) (f_2\!-\!1) f_2\!+\!2r f_2  f'_2+r^2 {f'_2}^2\big)\big)\!+\!f_1^2 r^2 \big({f'_1}^2 \big(2 (d\!-\!2) (d\!-\!1) f_2^2\!+\!r^2 {f'_2}^2\big)\nn \\
&+4r f_2  f'_1 \big((d-2) f_2+r f'_2\big) f''_1+4r^2 f_2^2  {f''_1}^2\big)\Big)\,,\\\label{use3}
R_{\m\n\a\b}R^{\m\n\a\b}&=\frac{1}{4r^4f_1^4}\Big(r^4 {f'_1}^4 f_2^2 +4 (d-2) f_1^4 (2 (d-3) (f_2-1)^2+r^2 {f'_2}^2)\nn \\
&-2 r^4 f_1 {f'_1}^2 f_2  ({f'_1} {f'_2}+2 f''_1 f_2)+f_1^2 r^2 \big({f'_1}^2 (4 (d-2) f_2^2+r^2 {f'_2}^2)\nn \\
&+4 r^2 f'_1 f''_1 f_2 f'_2+4 r^2 {f''_1}^2 f_2^2 \big)\Big)\,.
\end{align}
\end{subequations}
The sum of GH terms \eqref{EHGH} and  \eqref{RICGH} are given by
\begin{align}\label{use4}
\Big(g^{\a\b}&+\frac{\partial L^{Ricci}}{\partial R_{\a\m\b\n}} n_\mu n_\nu\Big) K_{\a\b}=\frac{f_2^\frac12}{2rf_1}\Big(2 (d-2) f_1+r f'_1\Big)+\frac{f_2^\frac12}{4f_1^3 r^3}
\Big(
(2 a_1+a_2) r^3 f_2 {f'_1}^3\nn \\
&-2 (d-2) f_1^3 \big(2 (d-3) (a_2+2 a_1 (d-2)) (f_2-1)+(4a_1(d-2)+a_2(d-1)) r f'_2\big)\nn \\
&-(2 a_1+a_2)r^3 f_1  f'_1 (f'_1 f'_2+2 f_2 f''_1)-2 (d-2) f_1^2 r \big(f'_1 \big((4 a_1+a_2) r f'_2-2 a_1 (d-3)\nn \\ 
&+(a_2+(6d-14) a_1) f_2\big)+(4 a_1+a_2) r f_2 f''_1\big)
\Big)\,.
\end{align}
The GH term \eqref{GHGB} has the following contribution
\begin{align}\label{use5}
 2 \Gg_{ab} {K}^{ab} &+\tfrac{1}{3}(K^3 - 3 K K_{ab} K^{ab} + 2 K_{ab} K^{bc} K_{c}{}^{a} ) =\nn \\
 &\frac{(d-3)(d-2)f_2^\frac12}{6f_1r^3}
 \Big(2 (d-4) f_1 (f_2-3)+3 r (f_2-1) f'_1\Big)\,.
\end{align}

\section{Action growth coefficients} \label{appendix B}
The coefficients of bulk action growth in equation \eqref{grbulk} are
\begin{align}\label{alphacoef}
&\a_1=(d-3) (d-2) q^2 \Big((d-2)^2 (d-1)+2 (d-3)\Lambda_0 \Big(-2 a_3 (d-4) (d-2)\nn \\
&+2 a_1 (d-4) (d-1) (2d-3)+a_2 (6+(d-4)d^2)\Big)\Big)\,,\nn \\ 
&\a_2=-(d-3) (d-2)^3 (d-1)^2 \Big((d-4) (4 a_1 (d-1)+a_2 d) q^2-a_3 (d-2) m^2\Big)\,,\nn \\
&\a_3=2 (d-3) (d-2)^4 (d-1)q^2 m \Big(a_3 (10-4d)+4 a_1 (d-4) (d-1)+a_2 (d-4) d\Big) \,,\nn \\
&\a_4=2 \Lambda_0 \Big((d-2)^2 (d-1)+2 d\Lambda_0 (a_3 (d-3) (d-2)+a_2 (d-1)+a_1 (d-1) d)  \Big)\,,\nn \\
&\a_5=(d-3) (d-2)^3 (d-1)  q^4\Big(a_3 (9d^2-45d+56 )-a_1 (d-4) (12d^2-45d+43)\nn \\
&-a_2 (-39+ 53d-23d^2+3 d^3)\Big)\,.
\end{align}
The coefficients of boundary action growth in equation \eqref{grbound} are
\begin{align}\label{betacoef}
\b_1&=2 (d\!-\!3) (d\!-\!2)^3 (d\!-\!1) \Big(3 a_1 (d\!-\!4) (31\!-
\!38d\!+\!11 d^2)\!+\!3 a_2 (9\!+\!4d\!-\!6d^2\!+\!d^3)\nn \\
&\!-\!a_3 (3d\!-\!7) (d^2\!+\!3d\!-\!16)\Big) q^4\,, \nn \\ 
\b_2&=4 (3d\!-\!7) \Lambda_0 \Big(3 (d\!-\!2)^2 (d\!-\!1)\!-\!2 a_3 (d\!-\!8) (d\!-\!3) (d\!-\!2) \Lambda_0\!+\!6 (d\!-\!1) d (a_2\!+\!a_1 d) \Lambda_0\Big)\,,\nn \\ 
\b_3&=2 (d\!-\!2)^3 (3d\!-\!7)\Big(\!-\!3 (d\!-\!2) (d\!-\!1)+4 (a_3 (d\!-\!4) (d\!-\!3)\!-\!3 (d\!-\!1)(a_2\!+\!d a_1)) \Lambda_0\Big)\,, \nn \\ 
\b_4&=\!-\!8 a_3 (d\!-\!4) (d\!-\!3) (d\!-\!2)^4 (d\!-\!1) (3d\!-\!7)\,, \nn \\ 
\b_5&=(d\!-\!2)^2 (d\!-\!1) (3d\!-\!7) m \Big(3 (d\!-\!2) (d\!-\!1)\!+\!4 (a_3 (d\!-\!3) (d\!-\!2)\!+\!3 (d\!-\!1)(a_2 \!+\!d a_1)) \Lambda_0\Big),\nn \\ 
\b_6&=4 a_3 (d\!-\!4) (d\!-\!3) (d\!-\!2)^4 (d\!-\!1) (3d\!-\!7) m\,,\nn \\
\b_7&=\!-\!2 (d\!-\!2) (3d\!-\!7) q^2 \Big(3 (d\!-\!2)^2 (d\!-\!1)\!+\!2 (d\!-\!3) (\!-\!2 a_3 (d\!-\!4) (d\!-\!2) (d\!+\!1)\nn \\
&\!+\!6 a_1 (d\!-\!4) (d\!-\!1) (3d\!-\!5)\!+\!3 a_2 (\!-\!6\!+\!22d\!-\!16d^2\!+\!3 d^3)) \Lambda_0)\,,\nn \\
\b_8&=\!-\!2 (d\!-\!3) (d\!-\!2)^3 (d\!-\!1) (3d\!-\!7) \Big(\!-\!3 \big(2 a_1 (d\!-\!4) (3d\!-\!5)
\!+\!a_2 (18\!-\!16d\!+\!3 d^2)\big) q^2\nn \\
&\!+\!a_3 (d\!-\!2) \big((d\!-\!1) m^2\!+\!2 (d\!-\!4) q^2\big)\Big)\,,\nn \\
\b_9&=\!-\!2 (d\!-\!3) (d\!-\!2)^3 (d\!-\!1) (3d\!-\!7) \Big(12 a_1 (d\!-\!4) (d\!-\!1)\!+\!3 a_2 (d\!-\!4) d\!-\!2 a_3 (d^2\!-\!7)\Big) m q^2\,.
\end{align}


\end{document}